\documentclass[a4paper]{article}
\usepackage[margin=2.5cm]{geometry} 
\usepackage{authblk}

\usepackage[utf8]{inputenc}
\usepackage[T1]{fontenc}
\usepackage{graphicx}
\usepackage{url}
\usepackage[backend=bibtex,style=numeric,sorting=none,doi=true,url=false,maxnames=5]{biblatex}
\begin{document}

\title{Towards Universal Visualisation of Emotional States for Information Systems\thanks{This is the postprint of a paper presented at the 32nd International Conference on Information Systems Development (ISD 2024) and published in its proceedings. DOI: 10.62036/ISD.2024.117}}

\author[1]{Michal R Wrobel}
\author[1]{Agnieszka Landowska}
\author[1]{Karolina Makuch}

\affil[1]{Faculty of Electronics, Telecommunications and Informatics, Gdansk University of Technology, ul. Narutowicza 11/12, Gdansk, 80-233, Poland}
\date{}

\maketitle

\begin{abstract}
The paper concerns affective information systems that represent and visualize human emotional states. The goal of the study was to find typical representations of discrete and dimensional emotion models in terms of color, size, speed, shape, and animation type. A total of 419 participants were asked about their preferences for emotion visualization. We found that color, speed, and size correlated with selected discrete emotion labels, while speed correlated with arousal in a dimensional model. This study is a first step towards defining a universal emotion representation for use in information systems. \par\vspace*{2mm}
\end{abstract}

\section{Introduction}
Emotional states are among the factors that moderate our engagement and focus. Affect influences the way people interact with information systems and the user experience. Automatic emotion recognition solutions have found applications in diverse fields~\cite{kolakowska2014emotion} such as opinion mining, market research, therapy, but also in communication and information systems.

The affective computing domain has developed not only the technologies to capture human affect, but also the solutions to express and influence affect.  
 When an emotional state is roughly quantified in a software environment, the result of the analysis should be effectively communicated to a potential human user to close the affective loop. Representations of emotional states should be easily perceived and understood by humans. There are a number of ways to communicate an emotional state within an information system, including labels, numbers, and visual representations. Visualization of emotional states has received some research attention in the last 10 years. A literature review conducted in 2022 by J. Wang et al. identified a number of approaches to visual encoding, although the majority of studies focused on visualizing emotions derived from music \cite{wang2022survey}. Visual representations can include avatar facial expressions, emojis, graphs/charts, graphs with labels, emojis or icons, colors, and shapes with/without animation.   

In this paper, we focus on the visual representation of affect within information systems in general, specifically exploring graphical and animation-based options available in an interface, assuming that few information systems involve a facial-expressive avatar. 
For example, a remote communication system proposed in a study by Ertay et al. uses emoji and colors as emotion representation, without exploration of color assignment to a particular affective state (happy is yellow, fearful is pink, and disgusted is green, surprise is black) \cite{enlay2021}. Another system for patient-physician communication uses colorful charts enhanced with emojis and labels (fear is yellow, sad is orange, surprise is turquoise)~\cite{Ma2023Aug}. A system for visualization of emotions in calligraphy uses colors only (calm is orange, anxious is aubergine, relaxed is green) \cite{calligraphy}. Another interesting study uses shape (cross-to-circle conversions), brightness, and size to represent pleasure-arousal-affinity dimensions of an emotional state \cite{Sanchez2013}. While in the study the usage of X-cross and circle is justified by human hand gestures (body language), the choice of the other representation methods is not discussed. However, the methods proposed in research papers for representing and visualizing emotional states in information systems are often contradictory, which shows the research gap that this article aims to fill. 

The aim of the study was to find typical techniques for the visual representation of emotions that could be interpreted in the same way by users of an information system. Both discrete and multidimensional models of emotions were considered. The research question addressed in this paper could be formulated as follows: \textit{Is it possible to define a typical representation for the visualization of a specific emotional state in terms of color, speed, size, shape, and animation}?

The paper describes a questionnaire-based study about visual representation of particular emotions and is organized as follows. The related works section provides a review of our starting point -- the emotion models, and emotion visualization techniques. The research methodology section describes the concept of the study, survey design decisions, and definition of variables. Then the results section provides findings on colors, size, speed, and animation types used in emotion visualization and is followed by discussion and conclusions sections.

\section{Related Work}
The studies that are most relevant to our research are those that deal with the representation and modeling of emotional states and those that aim to find a visualization for emotional states.

Two approaches to representing emotional states are commonly used: discrete (based on labels) and multidimensional~\cite{akolakowska2015}. The first of this group, such as Ekman's six basic emotions~\cite{ekman1971constants} or the PANAS model~\cite{watson1988development}, are more easily understood by humans and are therefore often used in surveys. On the other hand, multidimensional models, such as the Circumplex Model of Affect~\cite{russell1980circumplex} or the VAD (valence-arousal-dominance) ~\cite{mehrabian1996pleasure}, are more amenable to computer processing, and their mathematical representation allows them to be processed by algorithms for analyzing or synthesizing emotions. 
In emotion visualization, an emotional state described using the above models is mapped into a specific color, shape, animation, etc. to provide a sensory experience to a receiver of the state. The visualization aims at communication or enhancement of emotion-related information. In the study \cite{EMOPrints} authors proposed a concept of emotion-prints as an approach to visualize user emotional valence and arousal. They focused on real-time visualization of emotions to be used in standard GUI interfaces and formulated a thesis that ``the representation needs to be as intuitive as possible'' so that users could perceive and interpret visualization during real-time interaction. There are studies for color-based expressivity with a background in art \cite{art, expresivity}, however, those only partially transfer to information systems. There are some studies on how to visualize emotions in computer systems, but they focus on one feature (e.g. color) or one context (e.g. music). For example, so-called emotion scents based on color were proposed for valence-arousal plain representation for music-based emotions \cite{EmoScents}. Two studies explore diverse graphs for color-based representation of emotion state change in time \cite{timeseries2, timeseries}. Animated visualizations of Self-Assessment-Manikin (a well-established questionnaire for reporting emotional states in a dimensional model) were also explored for use as a self-reporting tool \cite{SAMAnim}. An interesting study on color \cite{AffColor} asked people to design palettes for states such as calm, playful, exciting, and disturbing. 

There are a few literature reviews on emotion visualization. One study by Pinilla et.al. explores visualization for virtual reality systems \cite{Pinilla}, highlighting object roundness, color in terms of brightness and saturation, and textures, applied to the positive-negative dimension of the emotional state model. Another study explores the visualization of sentiment from text, proposing a set of color-based tags and graphs, as well as quite interesting aggregation graphs representing anger and happiness \cite{Kucher}. A review study by Wang et.al. \cite{wang2022survey} examines a variety of visual encoding techniques for representing affect that have been used in research to date. The latter study is the most comprehensive one, but it does not provide a detailed methodology for its findings. For example, one of the figures shows emotional states' labels assigned to colors with a percentage, with no explanation of what the percentage represents.

To sum up, the literature review provided us with a number of visual representations, such as graphs/charts, graphs with labels, emojis or icons, colors, and shapes with/without animation to express emotion in information systems. With regard to the research question of the paper, we have not found a typical representation of an emotional state. There are many studies for color-emotion association, however we found the results contradictory. There are few or no studies that deal with other features such as element shape, size or animation speed and type and this is a research gap we intend to add to with our study.

\section{Research Methodology}

In order to answer the research question posed, a survey was prepared in the form of an interactive online questionnaire that was developed for the purpose of the study. Emotional states were visualized as a simple, configurable animation that could be used in different contexts like chats or dashboards. During the study participants configured the visualization using four parameters -- color, speed, size, and animation type.

\begin{figure*}[t]
\centering
\includegraphics[width=0.9\textwidth]{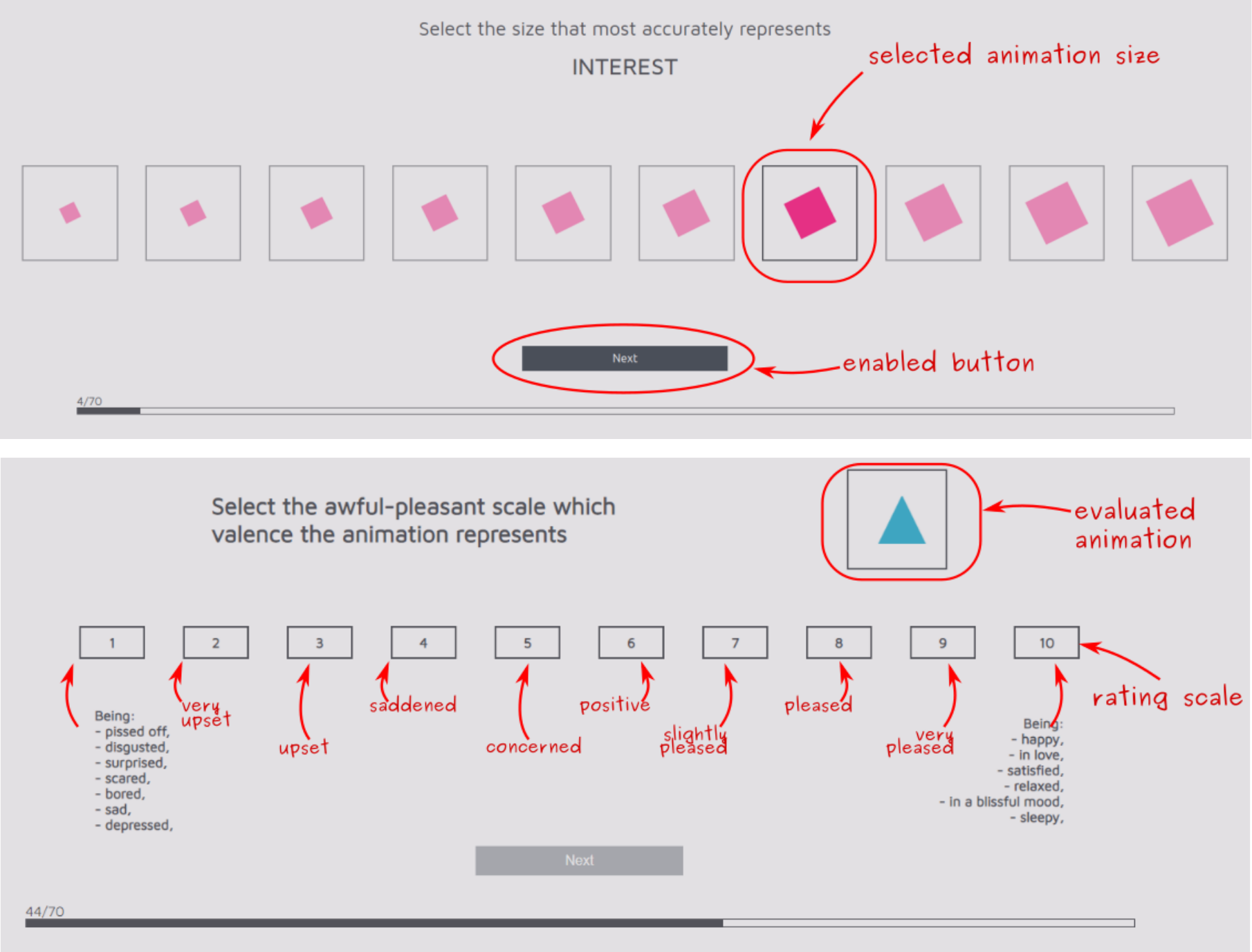}
\caption{Screenshots of the help screen for defining animation parameters (top) and specifying a values in the VAD model (bottom).}
\label{fig:app1}
\end{figure*}

At the beginning of the study, discrete emotion labels were selected for which a universal animation was to be developed. 
In the first step, the participant specified the values of the four animation parameters for each selected label, in subsequent steps on separate screens. This was done through a graphical interface, shown in the top screenshot in~Figure~\ref{fig:app1}. First the animation type was chosen, then the color, speed and finally the size.

In the second part of the study, for each animation created in the first step, the participant marked values for the three dimensions of the VAD model, namely valence, arousal and dominance. An animation was displayed and on three screens the participant determined the value for the awful-pleasant, calm-excited, and compilant-commanding scales, respectively, as shown in the bottom screenshot in Figure~\ref{fig:app1}. Sample labels for the extreme values of each scale were also displayed on the screen.

\subsection{Survey design decisions}
During the design phase of the survey application, decisions had to be made regarding the choice of emotion labels to be surveyed, as well as the set of colors and animations available. 

\textbf{Emotion labels:}
The choice of emotion labels to include in the study was determined by a trade-off between comprehensiveness and the amount of time participants were able to devote to the study. Given the expected high number of participants, models with a large number of labels, such as PANAS, PANAS-X, JES, or JAS~\cite{akolakowska2015}, were excluded. Ekman's model of basic emotions, on the other hand, seemed too limited, especially with regard to positive emotions. Therefore, Izard's model~\cite{izard1977differential} was chosen as it provides a balanced middle ground. Hence, the following labels were used in the study:  \textit{interest}, \textit{joy}, \textit{surprise}, \textit{sadness}, \textit{anger}, \textit{disgust}, \textit{contempt}, \textit{fear}, \textit{shyness}, and \textit{guilt}. 

\begin{figure}[t]
\centering
\includegraphics[width=0.99\textwidth]{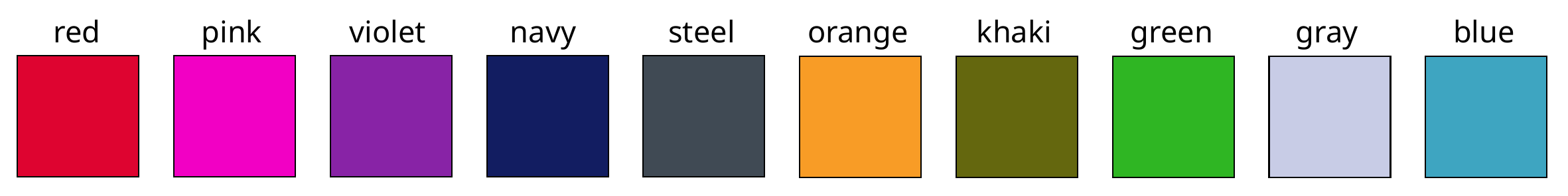}
\caption{Selected set of colors.}
\label{fig:colors}
\end{figure}

\textbf{Colors: }
Color selection was initially based on Plutchik's theory of emotions~\cite{plutchik1980general}. The goal was to present participants with a wide range of colors, significantly different from each other. The initial selection was then modified after a pilot study in which participants indicated little difference between some of the colors presented. Ultimately, the following colors were used in the study, as shown in Figure~\ref{fig:colors}: red, pink, violet, navy, steel, orange, khaki, green, gray, and blue.

\begin{figure*}[t]
\centering
\includegraphics[width=0.99\textwidth]{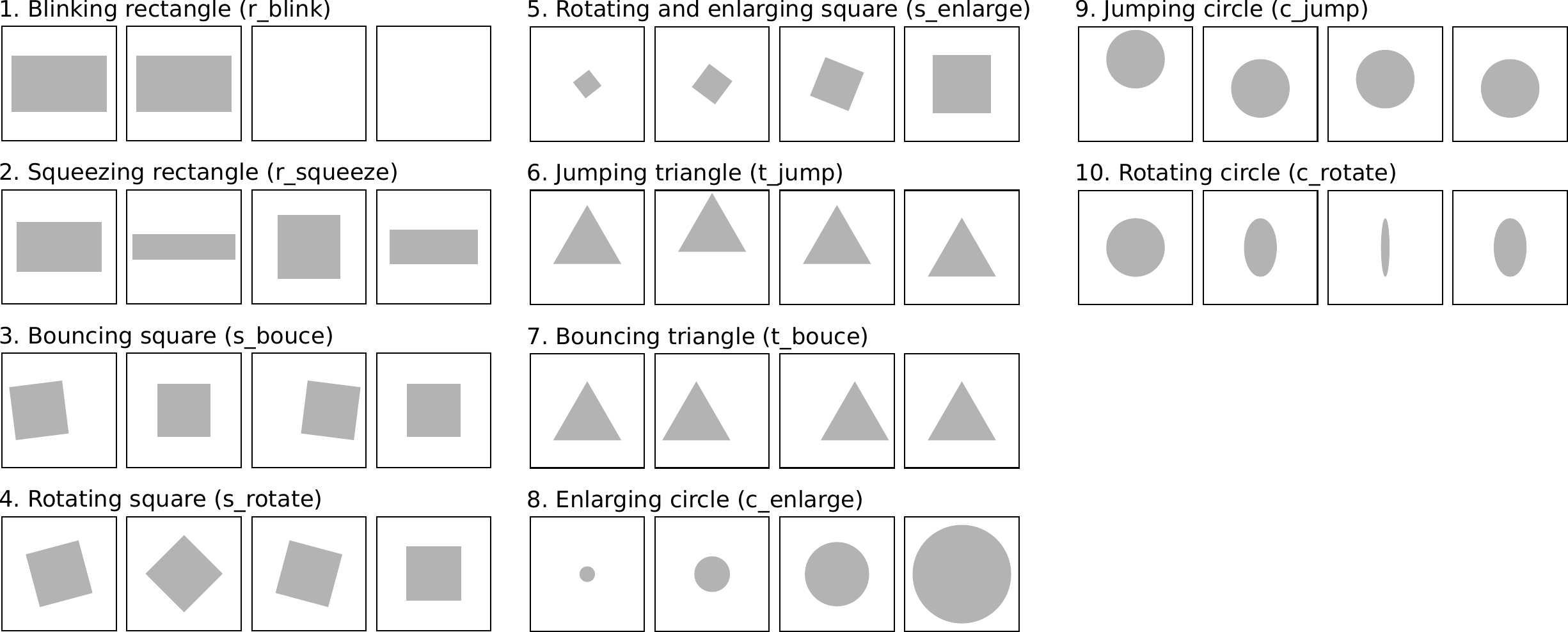}
\caption{Selected set of animations.}
\label{fig:animations}
\end{figure*}

\textbf{Animations: }
The main premise in selecting the animations was their simplicity, so that they could be used in a wide range of applications. For this reason, it was decided to use animated simple geometric figures such as rectangle, square, triangle, and circle. Each animation is looped, that is, when finished, it starts from the beginning. The first frame in Figure~\ref{fig:animations} shows the animation's initial state, the next at one-quarter, the next at half, and the last at three-quarters. 

Furthermore, each animation had two configurable parameters: speed and size, which could be set in the range 0.1 to 1. For the speed parameter, the lower the value, the slower the animation and the longer it took to animate from the first frame to the last. On the other hand, for the size parameter, the larger the value, the larger the animated figure. 

\subsection{Definition of Variables}
For the purpose of the analysis, a set of variables has been defined, which are listed in Table~\ref{tab:variables}. Four independent variables have been identified, four dependent and two confounding.

\begin{table*}[ht]
    \small
    \centering
    \caption{Definition of variables}
    \begin{tabular}{c|c|l|c|l} 
      Variable* & Short name & Full name & Scale type & Range
      \\
      \hline
      IV & VAL & Emotional state valence  & discrete** & \{1-10\} \\
      IV & ARO & Emotional state arousal & discrete** & \{1-10\} \\
      IV & DOM & Emotional state dominance & discrete** & \{1-10\} \\
      IV & EMO & Emotional state (label) & nominal & \{interest, joy, surprise, ...\} \\
      DV & TYPE & Animation shape and type & nominal & \{r\_blink, r\_squeeze, ... \} \\
      DV & COLOR & Color of animated shape & nominal & \{red, pink, violet, navy,, ...\} \\
      DV & SIZE & Size of animated shape & ordinal & \{0.1-1\} \\
      DV & SPEED & Speed of animation & ordinal & \{0.1-1\} \\
      CV & AGE & Age group of participant & nominal & \{pre-study, study, post-study\} \\
      CV & GENDER & Gender of participant & nominal & \{M-male, F-female, O-other\} \\
      \hline
    \end{tabular}

\end{table*}

\section{Results}

A total of 419 participants completed the survey, of whom 150 identified as female, 261 as male, 5 as other, and 3 did not specify. The age distribution of respondents is not even, with the vast majority between the ages of 18 and~32.

\subsection{Colors}

\begin{figure*}[htb]
\centering
\includegraphics[width=1\textwidth]{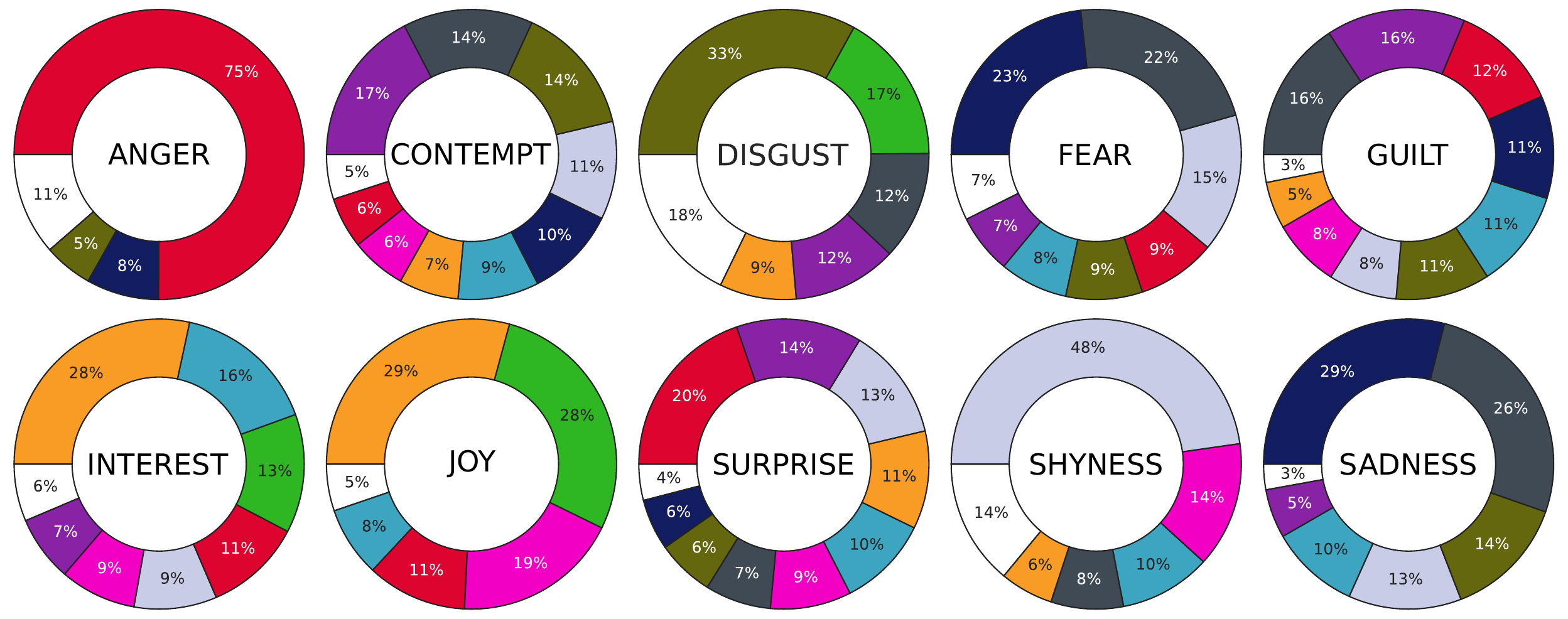}
\caption{Percentage of color indications for each emotion (white represents the sum of indications for colors with a share of less than 5\%).}
\label{fig:colors_result}
\end{figure*}

Based on the data collected, a summary of the colors chosen by participants for each emotion label was summarized and shown in Table~\ref{tab:colors_discrete}. Figure~\ref{fig:colors_result} presents a visualization of these results in the form of pie charts. For the purpose of clarity, less significant color indications, which accounted for less than 5\% of respondents, were combined and displayed as white in the figure.

The results reveal that only for one emotion, anger, one particular color was unambiguously indicated. In this case, more than 75\% of participants indicated the color red, with the share of other colors being well below 10\%. Another emotion with a single dominant color was shyness, where the color gray was selected by 47.7\% of participants. 

\begin{table*}[htb]
\small
    \centering
    \caption{Percentage share of specific colors for each emotion. }
    \begin{tabular}{|l|r|r|r|r|r|r|r|r|r|r|}
    \hline
\textbf{emotion} & \textbf{red} & \textbf{pink} & \textbf{violet} & \textbf{navy} & \textbf{steel} & \textbf{blue} & \textbf{gray} & \textbf{green} & \textbf{khaki} & \textbf{orange} \\
      \hline
anger & 75.1\% & 3.1\% & 2.6\% & 8.1\% & 2.1\% & 0.0\% & 0.7\% & 1.7\% & 5.5\% & 1.2\% \\ \hline
shyness & 1.2\% & 14.0\% & 3.6\% & 2.1\% & 8.1\% & 10.2\% & 47.7\% & 4.8\% & 2.4\% & 5.9\% \\ \hline
disgust & 2.1\% & 3.8\% & 11.6\% & 4.8\% & 12.1\% & 3.1\% & 4.0\% & 16.9\% & 33\% & 8.6\% \\ \hline
joy & 11.2\% & 18.5\% & 2.4\% & 0.2\% & 0.2\% & 7.8\% & 2.4\% & 28.0\% & 0.0\% & 29.2\% \\ \hline
sadness & 1.2\% & 0.7\% & 5.5\% & 29.0\% & 26.4\% & 10.0\% & 12.6\% & 0.2\% & 13.8\% & 0.7\% \\ \hline
interest & 10.9\% & 8.5\% & 7.3\% & 2.1\% & 2.6\% & 16.1\% & 9.2\% & 13.2\% & 1.7\% & 28.4\% \\ \hline
fear & 8.8\% & 1.7\% & 6.7\% & 23.3\% & 22.3\% & 7.6\% & 15.4\% & 2.1\% & 8.6\% & 3.6\% \\ \hline
surprise & 19.7\% & 9.0\% & 14.0\% & 5.7\% & 7.4\% & 10.2\% & 12.6\% & 4.0\% & 6.4\% & 10.9\% \\ \hline
contempt & 5.7\% & 6.2\% & 17.3\% & 10.2\% & 14.5\% & 9.0\% & 10.9\% & 5.0\% & 14.5\% & 6.7\% \\ \hline
guilt & 12.2\% & 7.6\% & 15.5\% & 11.5\% & 15.8\% & 11.0\% & 7.6\% & 3.1\% & 10.5\% & 5.3\% \\ 
      \hline
    \end{tabular}
    \label{tab:colors_discrete}
\end{table*}

Another group of emotions are those that are dominated by a set of similar colors. This group includes the emotion of disgust, for which khaki was the most frequently indicated color, followed by green and steel, colors of  moderate saturation and low value in the HSV color model. Likewise, for the emotions of sadness and fear, where the predominant colors were close to various shades of gray, such as navy, steel and gray. In both of these cases, the percentage of indications of the color khaki was also significant.

\begin{table}[t]
\centering
\small
\caption{Statistics for indicated speed and size values for each emotion. }
\begin{tabular}{|l|r|r|r|r|}
\hline
 & \multicolumn{2}{|c|}{\textbf{size}} & \multicolumn{2}{|c|}{\textbf{speed}} \\
 \hline
\textbf{emotion} & \multicolumn{1}{|c|}{\textbf{mean}} & \multicolumn{1}{|c|}{\textbf{std. dev.}} & \multicolumn{1}{|c|}{\textbf{mean}} & \multicolumn{1}{|c|}{\textbf{std. dev.}} \\
\hline
anger & 0.86 & 0.19 & 0.90 & 0.17 \\ \hline
contempt & 0.67 & 0.22 & 0.50 & 0.25 \\ \hline
disgust & 0.63 & 0.23 & 0.60 & 0.27 \\ \hline
fear & 0.65 & 0.30 & 0.69 & 0.31 \\ \hline
guilt & 0.59 & 0.27 & 0.44 & 0.26 \\ \hline
interest & 0.69 & 0.22 & 0.67 & 0.23 \\ \hline
joy & 0.80 & 0.19 & 0.82 & 0.19 \\ \hline
sadness & 0.56 & 0.29 & 0.27 & 0.20 \\ \hline
shyness & 0.35 & 0.26 & 0.30 & 0.23 \\ \hline
surprise & 0.72 & 0.22 & 0.69 & 0.26 \\ \hline
\end{tabular}
\label{tab:size_speed_discrete}
\end{table}

\begin{figure}[tbp]
\centering
\includegraphics[width=1\textwidth]{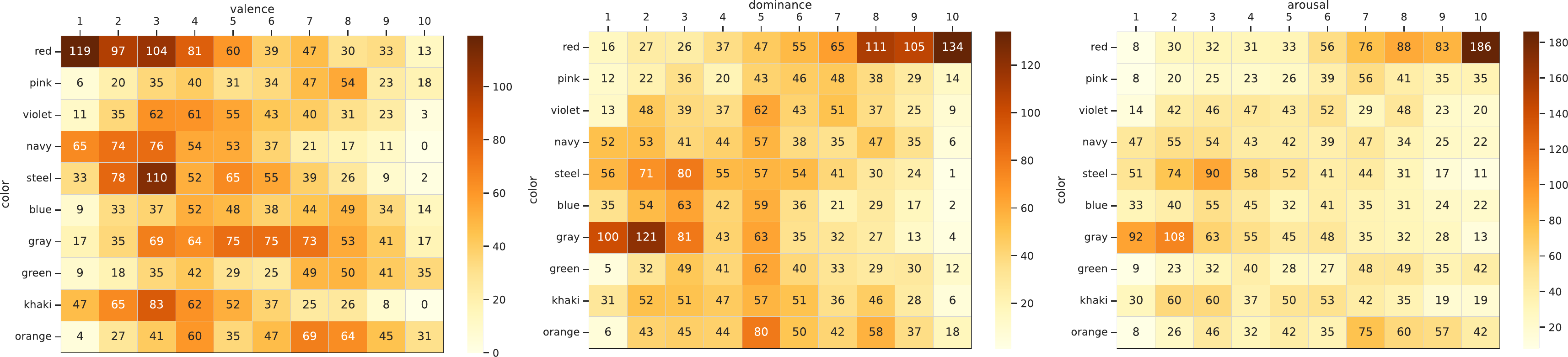}
\caption{Colors relationship with VAD dimensions.}
\label{fig:colors_vad}
\end{figure}

Finally, for positive emotions, interest and joy, orange was the dominant color.  For joy emotion, the next colors in terms of indications were characterized by high saturation, and these were green, pink and red. Meanwhile, for interest emotion, participants also frequently indicated blue and, to a lesser extent, green. 

For the other three emotions, contempt, surprise and guilt, no common features could be identified from the collected results that determined the choice of color. 

For the analysis of the colors chosen by the participants for the emotions in the three-dimensional model, the results are presented as a heatmap in the Figure~\ref{fig:colors_vad}. As with the discrete model, the only clear indication is for the color red. This color was most frequently selected for emotions of low valence and high arousal and dominance. This is consistent with the previous observation, as such values in the VAD model are most often identified with the emotion of anger. 

Of the remaining colors, it is possible to indicate a connection on all three axes of the VAD model for two colors in the shade of gray, i.e., steel and gray. These colors were most often indicated for low arousal and dominance and medium valence, with the darker shade -- steel -- more often chosen for low valence than the lighter one. In addition, the color orange can be associated with emotions of mid-high valence, mid-high arousal, and medium dominance.

The remaining colors, on the other hand, can be considered in terms of valence dimension. The colors navy and khaki were chosen more often by participants at low valence, and green at high. The other colors correspond to the average valence, with pink and orange skewed toward the higher values, and purple and navy toward the lower. 

For the arousal and dominance dimensions, in addition to the four colors described above, no significant connection can be found.

\begin{figure}[tbp]
\includegraphics[width=1\textwidth]{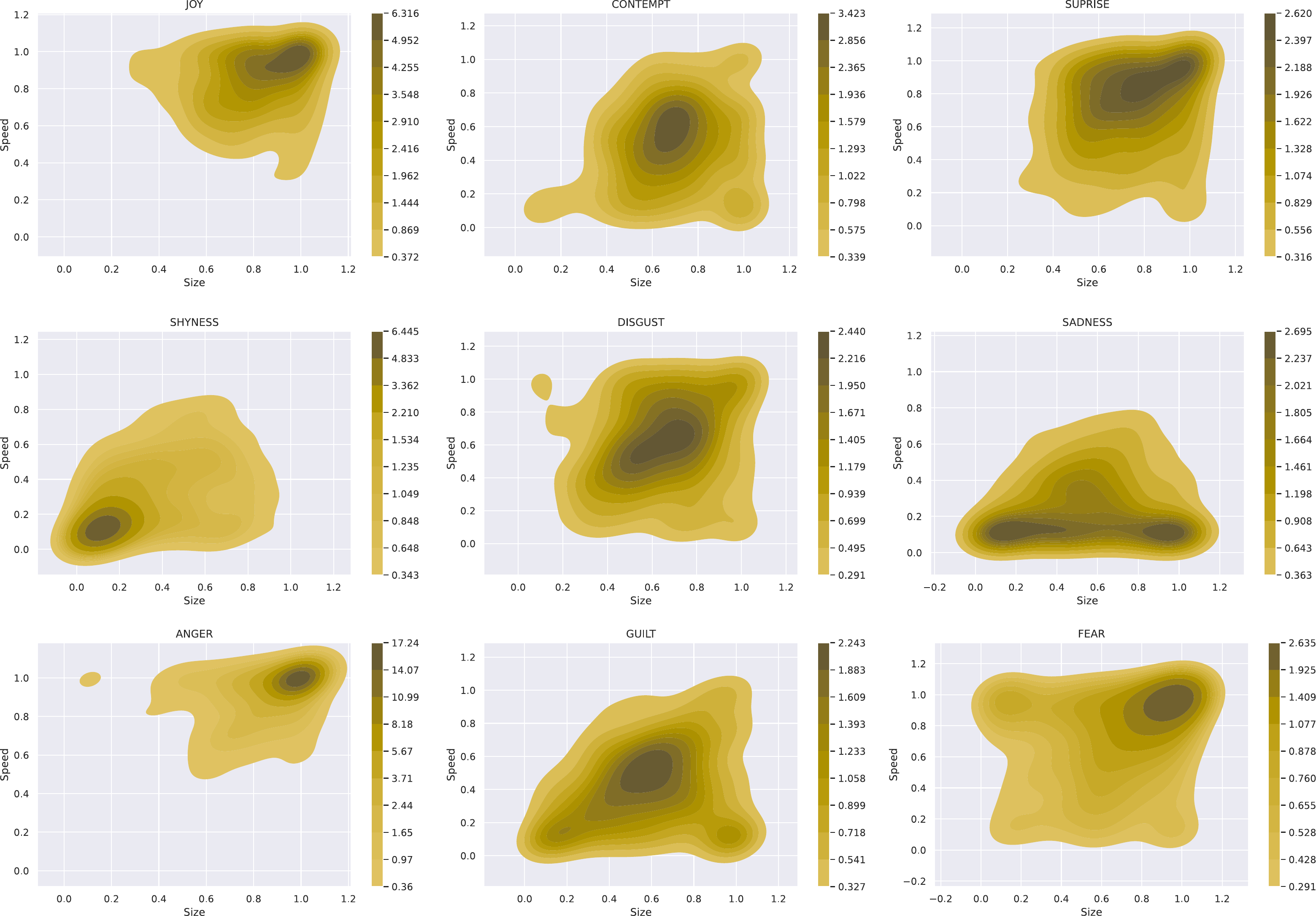}
\caption{Speed and size for each label.}
\label{fig:speed_size_result}
\end{figure}

\subsection{Size and speed}
For each discrete emotion label, the mean and standard deviation of the animation size and speed chosen by the participants were calculated, as shown in Table~\ref{tab:size_speed_discrete}. Variability ranges from 0.19 to 0.30 for size and 0.17 to 0.31 for speed. The emotions of anger and joy are characterized by the lowest variability, while the emotions of fear the highest. 

To provide a comprehensive illustration of the results, the values for size and velocity for each emotion are shown in the Kernel Distribution Estimation plot in Figure~\ref{fig:speed_size_result}. 

The emotion of anger is almost unambiguously associated by participants with fast and large animation. Responses are also dense for joy, and this emotion is likewise  associated with large and fast animations. On the other side is the emotion of shyness, for which low-speed and small-sized animations are indicated. 

Spearman's rank correlation coefficient was calculated to test the relationship between the individual dimensions of emotion represented in the multidimensional VAD model and the size and speed of the animation. The results are shown in Table~\ref{tab:size_speed_vad}.

\begin{table}[tb]
\centering
\small
\caption{Spearman's rank correlation coefficient between VAD dimensions and animation size and speed.}

\begin{tabular}{|l|r|r|}
\hline
& \multicolumn{1}{|c|}{\textbf{size}} & \multicolumn{1}{|c|}{\textbf{speed}} \\ \hline
\textbf{valence} & -0.073388 & -0.046321 \\ \hline
\textbf{arousal} & 0.313354 & 0.610777 \\ \hline
\textbf{dominance} & 0.333827 & 0.443621 \\ \hline
\end{tabular}
\label{tab:size_speed_vad}
\end{table}

The results reveal the highly moderate relationship between animation speed and arousal, with a correlation coefficient of 0.61. There is also a moderate correlation between dominance and speed, with a coefficient of 0.44. For speed, on the other hand, a weak correlation was found for both the arousal and dominance dimensions, with coefficients of 0.31 and 0.33, respectively. In contrast, the valence dimension is not related to either size or speed. 

\subsection{Animation type}
The survey results for the types of animations that participants chose for each emotion label are shown in Table~\ref{tab:animation_result}. In general, no strong correlation can be found between animation type and emotion. 

\begin{table*}[tb]
\fontsize{6.3}{8.8}\selectfont
\centering
\caption{Percentage share of specific animation for each emotion.}
\begin{tabular}{|l|r|r|r|r|r|r|r|r|r|r|}
\hline
\textbf{} & \multicolumn{10}{c|}{\textbf{animation}} \\ \hline
\textbf{emotion} & \multicolumn{1}{l|}{\textbf{r\_blink}} & \multicolumn{1}{l|}{\textbf{r\_squeeze}} & \multicolumn{1}{l|}{\textbf{s\_bouce}} & \multicolumn{1}{l|}{\textbf{s\_rotate}} & \multicolumn{1}{l|}{\textbf{s\_enlarge}} & \multicolumn{1}{l|}{\textbf{t\_jump}} & \multicolumn{1}{l|}{\textbf{t\_bouce}} & \multicolumn{1}{l|}{\textbf{c\_enlarge}} & \multicolumn{1}{l|}{\textbf{c\_jump}} & \multicolumn{1}{l|}{\textbf{c\_rotate}} \\
\hline
surprise & 28.5\% & 18.3\% & 4.0\% & 1.4\% & 8.1\% & 5.2\% & 4.0\% & 16.4\% & 4.5\% & 9.5\% \\ \hline
disgust & 9.7\% & 17.1\% & 11.6\% & 5.5\% & 4.8\% & 6.2\% & 22.3\% & 5.0\% & 2.1\% & 15.7\% \\ \hline
anger & 5.9\% & 21.1\% & 21.9\% & 7.4\% & 7.4\% & 4.3\% & 17.3\% & 9.5\% & 3.6\% & 1.7\% \\ \hline
joy & 0.7\% & 13.8\% & 18.1\% & 14\% & 10.9\% & 7.1\% & 5.0\% & 7.6\% & 21.9\% & 1.0\% \\ \hline
interest & 3.1\% & 11.8\% & 14.9\% & 9.7\% & 15.8\% & 3.8\% & 5.7\% & 21.5\% & 5.4\% & 8.3\% \\ \hline
shyness & 12.4\% & 3.6\% & 12.1\% & 4.0\% & 8.1\% & 6.7\% & 14\% & 10.2\% & 8.8\% & 20.2\% \\ \hline
sadness & 3.1\% & 9.0\% & 12.6\% & 5.5\% & 1.4\% & 11.6\% & 12.4\% & 7.6\% & 19.7\% & 17.1\% \\ \hline
fear & 15.7\% & 6.7\% & 11.6\% & 6.4\% & 9.7\% & 4.5\% & 16.6\% & 18.3\% & 5.2\% & 5.2\% \\ \hline
contempt & 6.4\% & 6.7\% & 13.5\% & 8.1\% & 4.5\% & 15.9\% & 13.1\% & 8.8\% & 6.9\% & 16.2\% \\ \hline
guilt & 6.0\% & 7.2\% & 12.9\% & 11.5\% & 4.5\% & 10.3\% & 9.3\% & 14.3\% & 12.2\% & 11.9\% \\ \hline
\textbf{total} & 9.0\% & 11.5\% & 13.3\% & 7.3\% & 7.5\% & 7.6\% & 12.0\% & 11.9\% & 9.0\% & 10.7\% \\ \hline
\end{tabular}
\label{tab:animation_result}
\end{table*}

\begin{figure}[tbp]
\centering
\includegraphics[width=1\textwidth]{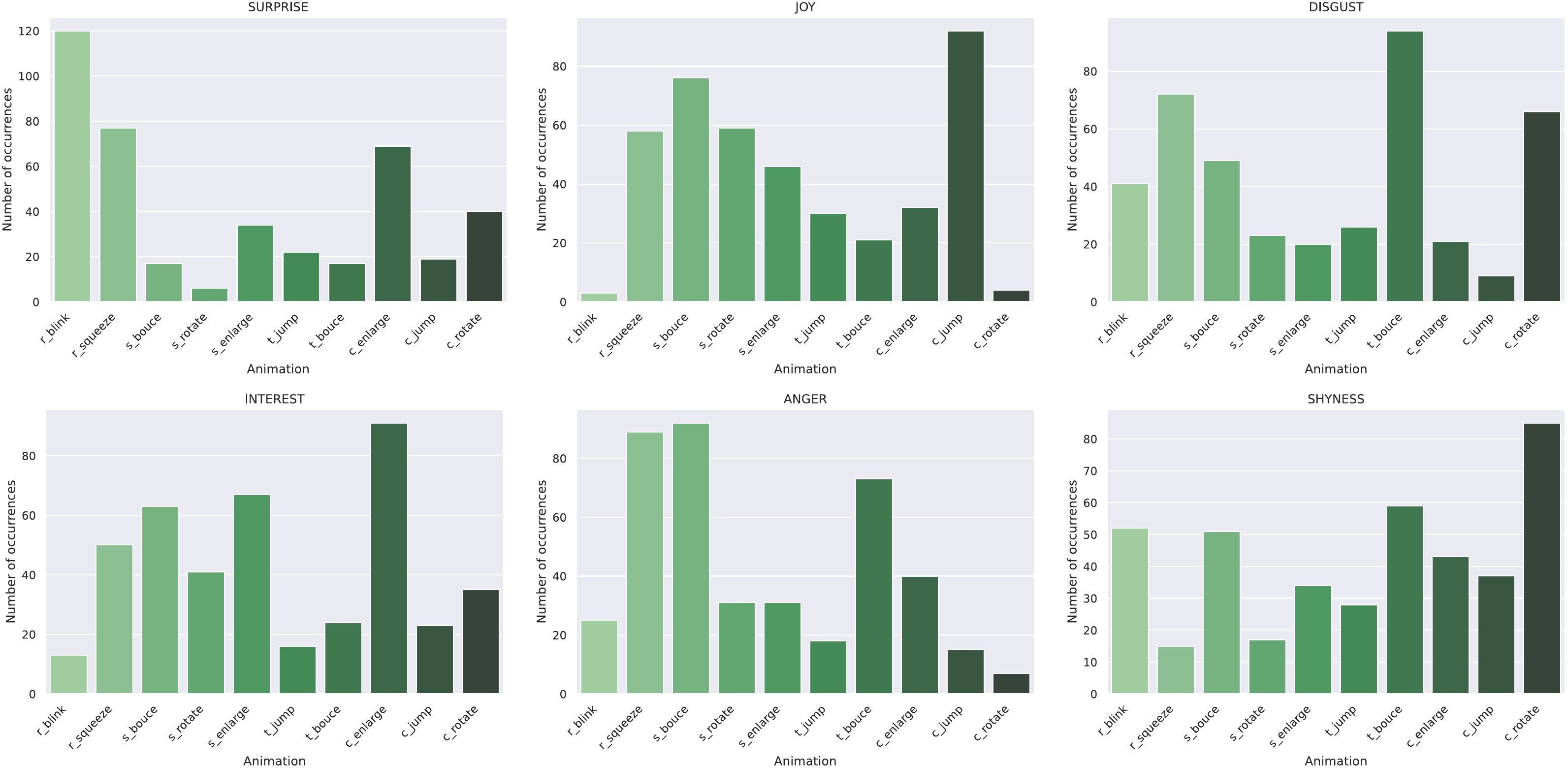}
\vspace{-0.8cm}
\caption{Animations types for six selected labels}
\label{fig:animation_result}
\end{figure}

However, for the four emotions where the number of respondents indicating the leading animation is greater than 20\%, some relationships can be seen. Animation ``r\_blink'' (blinking rectangle) may be associated with surprise with 28.5\% indications, while ``t\_bounce'' (bouncing triangle) with disgust (22.3\%)). Finally, ``r\_squeeze'' (squeezing rectangle) and ``c\_jump'' (jumping circle) may be linked with anger (21.1\%) and joy (21.1\%), respectively. 
A bar chart of the emotions most closely associated with the selected animation type is shown in Figure~\ref{fig:animation_result}.
However, for most of the other emotions and animations, there is no clear pattern or indication of any particular association. 

\begin{figure}[tbp]
\centering
\includegraphics[width=1\textwidth]{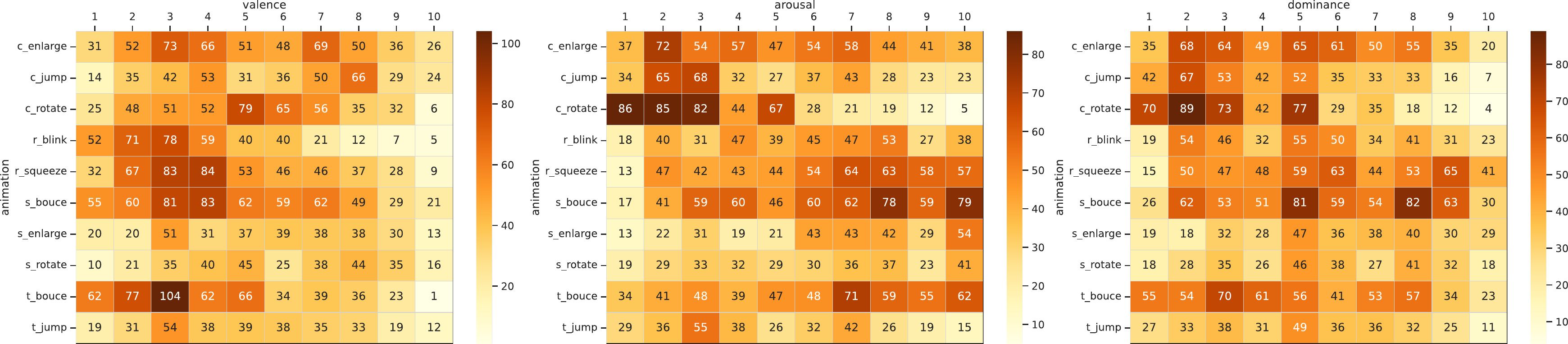}
\vspace{-0.6cm}
\caption{Animations relationship with VAD dimensions.}
\label{fig:animations_vad}
\vspace{-0.5cm}
\end{figure}

The survey data on the selection of participants for each dimension of the VAD model for each animation type is shown in the heatmap in Figure~\ref{fig:animations_vad}. Similar to the discrete emotion labels, there is no clear correlation between valence, arousal, or dominance and animation type, but some clues can be observed.

For the four animations, a link to the level of valence values can be observed. The ``t\_bouce'' and``r\_blink'' animations were most often reported by participants as representing a low valence value. For the former, this is consistent with the responses for the discrete model, where this animation was most often reported to represent disgust, anger, and fear -- all negative emotions. On the other hand, the ``c\_rotate'' and ``s\_bouce'' animations can be associated with the average values of this dimension.  

There are 5 animations for the arousal dimension that can be linked to its value. The ``c\_rotate'' animation was most often rated as the one with a low arousal value, and the same was true for ``c\_jump''. In contrast, the animations ``r\_squeeze'', ``s\_bouce'' and ``t\_bouce'' can be associated with high values of this dimension.  

Finally, two animations can be linked to the level of the dominance dimension -- ``c\_rotate'' for low values and ``s\_bouce'' for moderate. Therefore, for these two animations, clues can be identified in each dimension.

\subsection{Participants groups differences}

We assume that the attributes of size, color, and speed can be considered universal if statistical analyses do not reveal significant differences between subgroups defined by gender and age. Under these conditions, it can be concluded that such a representation of a particular emotion has a more universal character.  

For this purpose the nonparametric chi-square test for independence of variables in a contingency table was used. We assumed a threshold of 0.1 for testing the null hypothesis of a relationship of indications between groups. For p-values below the threshold, the null hypothesis should be rejected. On the other hand, the higher the p-value, the more the responses between groups are dependent on each other, i.e. the selected factors are more universal. In Tables~\ref{tab:chi_emotion_color} and~\ref{tab:chi_emotion_speed_size}, cells indicating acceptance of the null hypothesis are underlined. 

Table~\ref{tab:chi_emotion_color} shows the results of the chi-square test applied to the dependent variable color. The left side shows the results for the association of different colors with particular emotions, while the right side shows, conversely, the association of different emotions with specific colors. The latter shows which colors are more universally representative of a distinct emotion, thereby clarifying the color-emotion correlation across different contexts. Colors that are universal in this sense are: blue, dark blue, gray, khaki, orange, red, and violet. 

\begin{table}[tb]
\small
\centering
\caption{Chi-square tests results of independence of variables in a contingency table for color representation of emotions between genders and age groups.}
\begin{tabular}{|l|r|r|p{2cm}|l|r|r|}
\cline{1-3} \cline{5-7}
& \multicolumn{2}{|c|}{\textbf{p-value}} & & & \multicolumn{2}{|c|}{\textbf{p-value}} \\ \cline{2-3} \cline{6-7}
 \multicolumn{1}{|c|}{\textbf{emotion}} & \multicolumn{1}{|c|}{\textbf{gender}} & \multicolumn{1}{|c|}{\textbf{age}} & & \multicolumn{1}{|c|}{\textbf{color}} & \multicolumn{1}{|c|}{\textbf{gender}} & \multicolumn{1}{|c|}{\textbf{age}}\\ \cline{1-3} \cline{5-7}
anger & \underline{0.1827} & \underline{0.1451} & & amaranth & 0.0150 & \underline{0.4287} \\ \cline{1-3} \cline{5-7}
contempt & 0.0234 & \underline{0.1214}& & blue & \underline{0.2331} & \underline{0.4132} \\ \cline{1-3} \cline{5-7}
disgust & \underline{0.6102} & 0.0317 & & dark blue & \underline{0.6277} & \underline{0.3024} \\ \cline{1-3} \cline{5-7}
fear & 0.0584 & \underline{0.7867} & & gray & \underline{0.2852} & \underline{0.9965} \\ \cline{1-3} \cline{5-7}
guilt & \underline{0.4372} & \underline{0.8746} & & green & \underline{0.1441} & 0.0464 \\ \cline{1-3} \cline{5-7}
interest & 0.0338 & \underline{0.7587} & & khaki & \underline{0.2601} & \underline{0.8758} \\ \cline{1-3} \cline{5-7}
joy & 0.0348 & \underline{0.5211} & & orange & \underline{0.4926} & \underline{0.3167} \\ \cline{1-3} \cline{5-7}
sadness & 0.0237 & \underline{0.8428} & & red & \underline{0.1093} & \underline{0.3162} \\ \cline{1-3} \cline{5-7}
shyness & 0.0176 & 0.0293 & & steel & 0.0015 & \underline{0.8563} \\ \cline{1-3} \cline{5-7}
surprise & \underline{0.4220} & \underline{0.3983} & & violet & \underline{0.1418} & \underline{0.1554} \\ \cline{1-3} \cline{5-7}
\end{tabular}
\label{tab:chi_emotion_color}
\end{table}

\begin{table}[tb]
\small
\centering
\caption{Chi-square tests results of independence of variables in a contingency table for speed and size representation of emotions between genders and age groups.}
\begin{tabular}{|l|r|r|r|r|}
\hline
& \multicolumn{4}{|c|}{\textbf{p-value}} \\ \cline{2-5}
& \multicolumn{2}{|c|}{\textbf{gender}} & \multicolumn{2}{|c|}{\textbf{age}}\\ \cline{2-5}
 \multicolumn{1}{|c|}{\textbf{emotion}} & \multicolumn{1}{|c|}{\textbf{speed}} & \multicolumn{1}{|c|}{\textbf{size}} & \multicolumn{1}{|c|}{\textbf{speed}} & \multicolumn{1}{|c|}{\textbf{size}} \\ \hline
anger & \underline{0.7873} & \underline{0.2520} & 0.0003 & \underline{0.5579} \\ \hline
contempt & \underline{0.3651} & \underline{0.2169} & \underline{0.7811} & 0.0857 \\ \hline
disgust & \underline{0.8416} & \underline{0.1663} & \underline{0.7410} & \underline{0.6433} \\ \hline
fear & \underline{0.1040} & \underline{0.9583} & \underline{0.1492} & 0.0972 \\ \hline
guilt & \underline{0.7966} & \underline{0.1872} & \underline{0.9677} & \underline{0.6195} \\ \hline
interest & \underline{0.3215} & \underline{0.5431} & \underline{0.7618} & \underline{0.9127} \\ \hline
joy & \underline{0.3017} & \underline{0.3327} & \underline{0.9366} & \underline{0.5709} \\ \hline
sadness & \underline{0.1798} & \underline{0.8741} & \underline{0.5875} & \underline{0.1145} \\ \hline
shyness & 0.0952 & \underline{0.9094} & \underline{0.3795} & \underline{0.7334} \\ \hline
surprise & \underline{0.1350} & \underline{0.8140} & \underline{0.7959} & \underline{0.3342} \\ \hline
\end{tabular}
\label{tab:chi_emotion_speed_size}
\end{table}

Table~\ref{tab:chi_emotion_speed_size} shows the results of the chi-square test on the variables speed and size. The results show that these variables are universal between gender groups and between age groups for 6 of the 10 emotions. Emotions for which the p-value is less than 0.1 are: anger, contempt, fear, and shyness, and therefore their representation in terms of speed and size cannot be treated as universal. 

\section{Summary of results and discussion}
Analysis of the study results reveals several patterns of animation representation in both discrete and dimensional models of emotion. 

In the discrete emotion model, anger is clearly associated with the color red, as indicated by a strong consensus (75\%) and a p-value above the threshold, suggesting a universal association. Disgust is represented by khaki, green, and steel, all of which have moderate saturation and low value. Sadness and fear share a common representation in colors close to shades of gray, while positive emotions (joy and interest) are linked to orange and green. For dynamic representations, anger, joy, and fear are usually represented by animations that are fast and large, indicating their intensity. In contrast, shyness is represented by animations that are slow and small, reflecting its suppressed nature. Sadness follows this trend with low-speed animations, while contempt, guilt, and disgust are characterized by moderate speeds and sizes. Statistical tests conducted showed the universality of the relationship between animation speed and size and emotion labels. 

In the valence-arousal-dominance (VAD) dimensional model, red is typically associated with low valence, high arousal, and high dominance. Shades of gray such as steel and light gray are most often associated with low dominance, and the latter also with moderate valence. Furthermore, orange is associated with high valence. Statistical tests revealed a high to moderate correlation between arousal and speed. Surprisingly, although animation size is expected to correlate with dominance based on standard SAM representations~\cite{bradley1994measuring}, it shows only a weak relationship and is instead more aligned with speed.

With regard to color, our results explained why some findings for the color-emotion associations, from the previous studies we reported, were not the same. For example, sadness and fear both are associated with gray, and disgust can be represented with khaki, green, or steel, and it is not contradictory. In those terms, our study brings a new value of naming a set of colors to choose from while designing a palette for an information system interface, and shows which colors are unequivocal, and which ones might be ambiguous.

We also found relation between \textit{size} and \textit{shape} of an element and the emotional state. The data collected does not provide a clear link between \textit{animation type} and specific emotion, and this part of the study is inconclusive. 

\subsection{Validity threats} 
While these results suggest the potential for using color, shape, size, speed, and animation to visualize emotion, we note several limitations of our study.

First of all -- we studied a limited set of colors, shapes and animations. We also chose to cover a limited set of emotional states in the discrete model and a selected dimensional model. The limitations were imposed to keep the procedure simple and short, so that more participants would complete the questionnaire.
Another limitation is the use of an online questionnaire, which may introduce participation bias, as well as bias related to the display of colors on the various screens used by participants. The display of the same color may vary on different screens, potentially affecting user perception. While this issue could be mitigated in a controlled laboratory setting where all users use identical hardware, it would result in a smaller number of respondents. On the other hand, since the goal is to propose a universal visualization, it is reasonable to collect data from respondents using different types of devices, as they would do using the actual information systems.
Another issue is that we have only tested a limited range of emotion representations in an abstract context, and translating these findings to actual interfaces may reveal different phenomena.

Finally, even within our limitations, our results do not provide precise guidance for every emotional state.
However, the findings do offer some further directions to pursue.

\section{Conclusion}
There is a number of studies that confirm the influence of color on our emotions and mood, and this psychological effect is widely used in cinematography. In a famous book entitled ``If it's purple, someone's gonna die: the power of color in visual storytelling'' Patti Bellantoni explores color as a means for expression of scene or character mood~\cite{cinema}. Cinematography obviously differs from the design of information systems, and the color palette, as well as other means that might be applied, differ. The emotions felt while interacting with systems are more subtle, and the means for emotion visualization are less expressive. Still, there is a need to find visual techniques that would allow information systems to envision or evoke an emotional state. Although analysis of literature provided us with a number of studies, the results were frequently contradictory, and therefore we decided to perform our study, looking forward to contributing towards finding universal (or at least typical) emotion visualization.

Given the diversity of human perception and experience, it may be impossible to find a universal emotion visualization that would be interpreted in the same way by all people. However, the discussion of the mapping of a color or shape to an emotional state is crucial for the interface design of affective information systems. With our study, we have identified some interesting directions for further research. These include the exploration of other colors, shapes, animations, and emotional states, as well as personalized palettes for emotion visualization. In further research, we also plan to collect and compare components suitable for creating emotion visualizations, as well as thoroughly examine entire interfaces designed for this purpose.

In addition, we would like to make an important remark with regard to our research. The visualization of emotional states could offer some benefits in many application areas, such as empathic communication, artistic expression, the detection of stress or depression, etc. However, despite good intentions, technologies that detect and visualize human affect can also pose significant dangers by exposing people's inner states to others, and should therefore be used with care and responsibility.


\printbibliography

@article{wang2022survey,
	author = {Wang, Jiaqi and Gui, Tianyi and Cheng, Mingzhi and Wu, Xuan and Ruan, Ruolin and Du, Meng},
	title = {{A survey on emotional visualization and visual analysis}},
	journal = {Journal of Visualization},
	volume = {26},
	number = {1},
	pages = {177--198},
	year = {2023},
	month = feb,
	issn = {1875-8975},
	publisher = {Springer Berlin Heidelberg},
	doi = {10.1007/s12650-022-00872-5}
}

@incollection{akolakowska2015,
    title = {Modeling emotions for affect-aware applications},
	year = {2015},
	author = {Ko{\l}akowska, Agata and Landowska, Agnieszka and Szwoch, Mariusz and Szwoch, Wioleta and Wr{\'o}bel, Micha{\l}},
	booktitle = {Information Systems Development and Applications},
    publisher = {Faculty of Management, University of Gdańsk, Polanf}
}

@article{ekman1971constants,
  title={Constants across cultures in the face and emotion.},
  author={Ekman, Paul and Friesen, Wallace V},
  journal={Journal of personality and social psychology},
  volume={17},
  number={2},
  pages={124},
  year={1971},
  publisher={American Psychological Association}
}

@article{watson1988development,
  title={Development and validation of brief measures of positive and negative affect: the PANAS scales.},
  author={Watson, David and Clark, Lee Anna and Tellegen, Auke},
  journal={Journal of personality and social psychology},
  volume={54},
  number={6},
  pages={1063},
  year={1988},
  publisher={American Psychological Association}
}

@article{russell1980circumplex,
  title={A circumplex model of affect.},
  author={Russell, James A},
  journal={Journal of personality and social psychology},
  volume={39},
  number={6},
  pages={1161},
  year={1980},
  publisher={American Psychological Association}
}

@article{mehrabian1996pleasure,
  title={Pleasure-arousal-dominance: A general framework for describing and measuring individual differences in temperament},
  author={Mehrabian, Albert},
  journal={Current Psychology},
  volume={14},
  pages={261--292},
  year={1996},
  publisher={Springer}
}

@article{izard1977differential,
  title={Differential emotions theory},
  author={Izard, Carroll E and Izard, Carroll E},
  journal={Human emotions},
  pages={43--66},
  year={1977},
  publisher={Springer}
}

@incollection{plutchik1980general,
  title={A general psychoevolutionary theory of emotion},
  author={Plutchik, Robert},
  booktitle={Theories of emotion},
  pages={3--33},
  year={1980},
  publisher={Elsevier}
}

@inproceedings{enlay2021,
author = {Ertay, Eyl\"{u}l and Huang, Hao and Sarsenbayeva, Zhanna and Dingler, Tilman},
title = {Challenges of Emotion Detection Using Facial Expressions and Emotion Visualisation in Remote Communication},
year = {2021},
isbn = {9781450384612},
publisher = {Association for Computing Machinery},
address = {New York, NY, USA},
url = {https://doi.org/10.1145/3460418.3479341},
doi = {10.1145/3460418.3479341},
booktitle = {Adjunct Proceedings of the 2021 ACM International Joint Conference on Pervasive and Ubiquitous Computing and Proceedings of the 2021 ACM International Symposium on Wearable Computers},
pages = {230–236},
numpages = {7},
location = {Virtual, USA},
series = {UbiComp/ISWC '21 Adjunct}
}

@article{Ma2023Aug,
	author = {Ma, Hua and Sun, Xu and Lawson, Glyn and Wang, Qingfeng and Zhang, Yaorun},
	title = {{Visualising emotion in support of patient-physician communication: an empirical study}},
	journal = {Behaviour {\&} Information Technology},
	year = {2023},
	month = aug,
	publisher = {Taylor {\&} Francis},
	doi = {10.1080/0144929X.2022.2097954}
}

@article{calligraphy,
AUTHOR = {Wang, Chao-Ming and Chen, Yu-Chen},
TITLE = {Design of an Interactive Mind Calligraphy System by Affective Computing and Visualization Techniques for Real-Time Reflections of the Writer’s Emotions},
JOURNAL = {Sensors},
VOLUME = {20},
YEAR = {2020},
NUMBER = {20},
ARTICLE-NUMBER = {5741},
URL = {https://www.mdpi.com/1424-8220/20/20/5741},
PubMedID = {33050320},
ISSN = {1424-8220},
DOI = {10.3390/s20205741}
}

@INPROCEEDINGS{Sanchez2013,
  author={Garcia Sanchez, Jesus A. and Shibata, Atsushi and Ohnishi, Kazuhiro and Dong, Fangyan and Hirota, Kaoru},
  booktitle={2014 International Conference on Humanoid, Nanotechnology, Information Technology, Communication and Control, Environment and Management (HNICEM)}, 
  title={Visualization method of emotion information for long distance interaction}, 
  year={2014},
  volume={},
  number={},
  pages={1-6},
  doi={10.1109/HNICEM.2014.7016201}}

@article{bradley1994measuring,
  title={Measuring emotion: the self-assessment manikin and the semantic differential},
  author={Bradley, Margaret M and Lang, Peter J},
  journal={Journal of behavior therapy and experimental psychiatry},
  volume={25},
  number={1},
  pages={49--59},
  year={1994},
  publisher={Elsevier},
  doi={10.1016/0005-7916(94)90063-9}
}

@inproceedings{EMOPrints,
author = {Cernea, Daniel and Weber, Christopher and Ebert, Achim and Kerren, Andreas},
year = {2015},
month = {02},
pages = {},
title = {Emotion-Prints: Interaction-Driven Emotion Visualization on Multi-Touch Interfaces},
volume = {9397},
journal = {Proceedings of SPIE - The International Society for Optical Engineering},
doi = {10.1117/12.2076473}
}

@inproceedings{expresivity,
author = {St\r{a}hl, Anna and Sundstr\"{o}m, Petra and H\"{o}\"{o}k, Kristina},
title = {A foundation for emotional expressivity},
year = {2005},
isbn = {159593250X},
publisher = {AIGA: American Institute of Graphic Arts},
address = {New York, NY, USA},
booktitle = {Proceedings of the 2005 Conference on Designing for User EXperience},
pages = {33–es},
location = {, San Francisco, California, USA, },
series = {DUX '05}
}

@article{art,
author = {Bialoskorski, Leticia and Westerink, Joyce and van den Broek, Egon L.},
year = {2009},
month = {08},
pages = {173-191},
title = {Mood Swings: Design and evaluation of affective interactive art},
volume = {15},
journal = {The New Review of Hypermedia and Multimedia},
doi = {10.1080/13614560903131898}
}

@INPROCEEDINGS{EmoScents,
  author={Dharmapriya, Januka and Dayarathne, Lahiru and Diasena, Tikiri and Arunathilake, Shiromi and Kodikara, Nihal and Wijesekera, Primal},
  booktitle={2021 International Conference on Electronics, Information, and Communication (ICEIC)}, 
  title={Music Emotion Visualization through Colour}, 
  year={2021},
  volume={},
  number={},
  pages={1-6},
  doi={10.1109/ICEIC51217.2021.9369788}}

@inproceedings{timeseries,
author = {Sheidin, Julia and Lanir, Joel and Kuflik, Tsvi},
year = {2019},
month = {09},
pages = {1-9},
title = {A comparative evaluation of techniques for time series visualizations of emotions},
isbn = {978-1-4503-7190-2},
journal = {CHItaly '19: Proceedings of the 13th Biannual Conference of the Italian SIGCHI Chapter: Designing the next interaction},
doi = {10.1145/3351995.3352054}
}

@inproceedings{SAMAnim,
author = {Sonderegger, Andreas and Heyden, Klaus and Chavaillaz, Alain and Sauer, Juergen},
title = {AniSAM \& AniAvatar: Animated Visualizations of Affective States},
year = {2016},
isbn = {9781450333627},
publisher = {Association for Computing Machinery},
address = {New York, NY, USA},
url = {https://doi.org/10.1145/2858036.2858365},
doi = {10.1145/2858036.2858365},
booktitle = {Proceedings of the 2016 CHI Conference on Human Factors in Computing Systems},
pages = {4828–4837},
numpages = {10},
location = {San Jose, California, USA},
series = {CHI '16}
}

@inproceedings{AffColor,
author = {Bartram, Lyn and Patra, Abhisekh and Stone, Maureen},
title = {Affective Color in Visualization},
year = {2017},
isbn = {9781450346559},
publisher = {Association for Computing Machinery},
address = {New York, NY, USA},
url = {https://doi.org/10.1145/3025453.3026041},
doi = {10.1145/3025453.3026041},
booktitle = {Proceedings of the 2017 CHI Conference on Human Factors in Computing Systems},
pages = {1364–1374},
numpages = {11},
location = {Denver, Colorado, USA},
series = {CHI '17}
}

@INPROCEEDINGS{timeseries2,
  author={Bista, Sanat Kumar and Nepal, Surya and Paris, Cécile},
  booktitle={2014 IEEE International Congress on Big Data}, 
  title={Multifaceted Visualisation of Annotated Social Media Data}, 
  year={2014},
  volume={},
  number={},
  pages={699-706},
  doi={10.1109/BigData.Congress.2014.103}}

@article{Pinilla,
author = {Pinilla, Andres and Garcia, Jaime and Raffe, William and Voigt-Antons, Jan-Niklas and Spang, Robert and Möller, Sebastian},
year = {2021},
volume={2},
pages = {},
title = {Emotion visualization in Virtual Reality: An integrative review},
journal={Frontiers in Virtual Reality},
DOI={10.3389/frvir.2021.630731}
}

@article{Kucher,
author = {Kostiantyn Kucher and Teri Schamp-Bjerede and Andreas Kerren and Carita Paradis and Magnus Sahlgren},
title ={Visual analysis of online social media to open up the investigation of stance phenomena},
journal = {Information Visualization},
volume = {15},
number = {2},
pages = {93-116},
year = {2016},
doi = {10.1177/1473871615575079},
}

@article{kolakowska2014emotion,
  title={Emotion recognition and its applications},
  author={Ko{\l}akowska, Agata and Landowska, Agnieszka and Szwoch, Mariusz and Szwoch, Wioleta and Wrobel, Michal R},
  journal={Human-computer systems interaction: Backgrounds and applications 3},
  pages={51--62},
  year={2014},
  publisher={Springer}
}

@book{cinema,
	author = {Bellantoni, Patti},
	title = {{If It's Purple, Someone's Gonna Die: The Power of Color in Visual Storytelling}},
	year = {2005},
	isbn = {978-0-24080688-4},
	publisher = {Routledge}
}


\end{document}